\documentclass{article}
\usepackage{spconf,amsmath,graphicx}
\usepackage{multirow}
\usepackage{graphics}
\usepackage{threeparttable}
\usepackage{amsfonts}
\usepackage{amsmath}
\usepackage[shortlabels]{enumitem}
\setlist[enumerate]{nosep}

\title{X-SepFormer: End-to-end Speaker Extraction Network with Explicit Optimization on Speaker Confusion}
\name{Kai Liu, Ziqing Du, Xucheng Wan, Huan Zhou}
\address{AI Application Research Center, Huawei Technologies Co., Ltd., Shenzhen, China}
\begin{document}
%
\maketitle
\begin{abstract}
Target speech extraction (TSE) systems are designed to extract target speech from a multi-talker mixture. The popular training objective for most prior TSE networks is to enhance reconstruction performance of extracted speech waveform. However, it has been reported that a TSE system delivers high reconstruction performance may still suffer low-quality experience problems in practice. One such experience problem is wrong speaker extraction (called speaker confusion, SC), which leads to strong negative experience and hampers effective conversations. To mitigate the imperative SC issue, we reformulate the training objective and propose two novel loss schemes that explore the metric of reconstruction improvement performance defined at small chunk-level and leverage the metric associated distribution information. Both loss schemes aim to encourage a TSE network to pay attention to those SC chunks based on the said distribution information. On this basis, we present X-SepFormer, an end-to-end TSE model with proposed loss schemes and a backbone of SepFormer. Experimental results on the benchmark WSJ0-2mix dataset validate the effectiveness of our proposals, showing consistent improvements on SC errors (by 14.8\% relative). Moreover, with SI-SDRi of 19.4 dB and PESQ of 3.81, our best system significantly outperforms the current SOTA systems and offers the top TSE results reported till date on the WSJ0-2mix.
\end{abstract}

\section{Introduction}
In real multi-talker communication scenario, it is common that the speech from a speaker of interest is interfered with others. Such a corrupted speech mixture may bother human listeners with partially intelligible speech or make adverse impacts for downstream applications (such as speech recognition and speaker diarization). To get the desired speech from such a complex auditory scene, two deep neural network based strategies, speech separation (SS) and target speaker extraction (TSE) have been extensively studied recently with distinctive objectives.

The objective of SS is to recover all individual speakers from the mixture signal. Most SS algorithms are designed to estimate time-frequency separation masks or filters and optimized to minimize the signal reconstruction loss, which is proven generally having a good correlation with speech quality. On the most widely used dataset WSJ0-2mix \cite{hershey2016deep}, for example, the SepFormer can achieve performance that is close to the upper bound of the dataset (22.3 vs. 23.1 dB) \cite{lutati22}, in terms of scale-invariant signal-to-distortion ratio improvement (SI-SDRi) \cite{roux19, isik2016single}.

Alternatively, TSE aims to a more practical objective by extracting the speech from the target speaker while ignoring others. As a special case of SS, TSE performs both separation and speaker identification, with the aid of reference utterance from the target speaker. However, to extend a well-performed SS system for TSE task is non-trivial and the main challenge lies in how to properly extract and leverage the speaker information to benefit the extracted speech quality. 

To extract proper speaker information, most prior arts \cite{wang19h,chenglin19,li20p,katerina19,eskimez22,wang21x} directly utilize the well designed speaker embeddings, which are extracted from pre-trained networks designed for speaker verification task. Alternatively, motivated to produce helpful speaker information for speech separation, some studies \cite{ge2020spex+,marc20,katerina19} introduce an auxiliary neural network, which is jointly optimized with the main speech extraction network. 

Another line of research investigates proper fusion strategy to integrate the speaker information with the speech mixture. Many speaker-speech fusion schemes have been proposed for effectively guiding the separator, such as concatenation \cite{jian19,chenglin19,li20p}, multiplication \cite{katerina19,elminshawi22}, scaling adaptation \cite{zeghidour21,wang21x} and attention-based adaptation \cite{jiangyu21,wang21,tsubasa19}. 


Despite all endeavors, the state-of-the-art (SOTA) TSE system fails to outperform a top SS system. Indeed, on the benchmark WSJ0-2mix, the performance of the best TSE model is 18.6 dB, much lower than the record of SepFormer (22.3 dB). What's more, it has been observed that the TSE extracted speech suffers low quality problem as well. These could be negatively experienced as speaker identification error, disruptive (over-suppressed or under-suppressed) speech and non-silent output for inactive speaker, as reported before in \cite{zifeng22,elminshawi22,eskimez22,zexu22,zhang20m,borsdorf21,marc22}. Among them, the most unpleasant experience may be the speaker identification error (also called speaker confusion, SC), in which the speech from interfering speaker is extracted while desired target speech is filtered out. Losing target speech, even occasionally for a short period, is highly undesired and hampers effective conversations.


However, we find only a few studies \cite{zhang20m,zifeng22,elminshawi22} that relate to the SC problem (details would be introduced in Section 2). As a short summary, a common solution is to apply a post-processing during inference to identify and correct SC-related TSE output. The main shortcoming of the solution is that it incurs additional computational cost; and by nature, it is inept to handle more practical case where the extracted output is partially destroyed by SC. An alternative solution is to modify the loss function,which however has unrealistic requirement of information from the interfering speaker (like speech input or reference utterance). 

In this paper, we propose a new TSE model, referred as X-SepFormer. It extends SepFormer for TSE task, with twofold objectives: i) to explicitly address the critical SC problem during the training stage without additional requirement about the interfering speaker; ii) to bridge the performance gap between TSE and SOTA SS, say on the dataset WSJ0-2mix.

The rest of the paper is organized as follows. SC-related TSE works are briefly reviewed in Section 2. Section 3 describes our proposals. Experimental results are reported and analyzed in Section 4. Finally we conclude the paper in Section 5. 

\section{Related Works}
Given target speech $x_s$, interfering speech $x_i$ and extracted output $\hat x$, various methods to measure SC or solve SC were proposed in a few SC-related literatures.

In X-TasNet \cite{zhang20m}, the speech extraction quality was measured using metric of negative SI-SNRi rate (NSR), which was proven to be a good approximation to the subjective speaker error rate. By employing combined SI-SNR losses on both the target and interfering speaker, $\text{SI-SNR}(\hat x, x_{s})+\text{SI-SNR}(\hat x, x_{i})$, and training scheme, X-TaSNet showed enhanced quality on both SI-SNRi and NSR. Similar to the idea, \cite{elminshawi22} also found that very negative $\Delta \text{SI-SDR}$ corresponds to wrong speaker selection; and proposed a quantitative analysis, $\text{SI-SDR}(\hat x, x_{i}) - \text{SI-SDR}(\hat x, x_{s})\geq 8 dB$, to measure how often the interferer is selected. 

The study \cite{marc22} investigated inactive speaker issue in TSE. It proposed a post active/inactive speaker detection method based on cosine-similarity-based speaker verification. At test time, inactive speaker or extraction failure is detected if the cosine similarity between embeddings is lower than the threshold, $\mathcal{C}(e^{\hat x}, e^{x_{s}}) \leq \eta^{Cos}$, and according extracted output is set 
to zeros. Very recently, considering the importance of distinctive speaker embeddings for end-to-end speaker extraction, \cite{zifeng22} proposed to solve the SC problem in two stages. In the training phase, it adopts combined reconstruction loss and the metric learning loss to improve the distinguishability of embeddings; while for inference, a post-filtering strategy is designed to revise the wrong results, identified by measuring the $L_2$ distances between estimated output and two enrollment from target and interfering speaker, respectively.


\begin{figure}
\centering
\includegraphics[width=0.50\textwidth]{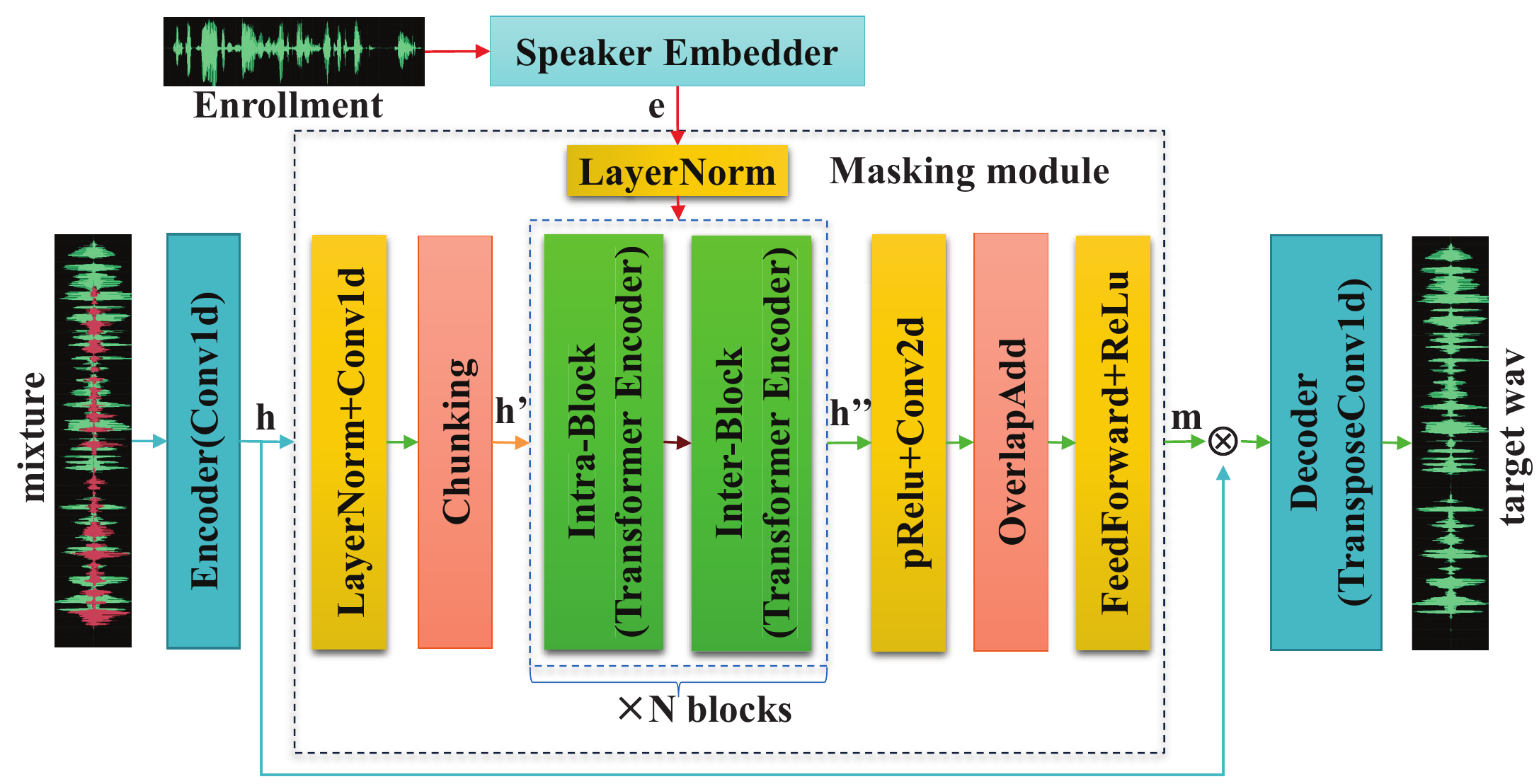}
\caption{The architecture of proposed X-SepFormer. }
\label{fig1}
\end{figure}

\section{proposed method}
In this section, we approach the problem described in Section 1 from two different perspectives. First, we describe X-SepFormer, an end-to-end TSE model that is built on SepFormer. Next, For the first time, we raise the segmental SC issue. To address the issue in X-SepFormer, we design two new loss schemes, which are also appealing that can be applied to any TSE model. 
\subsection{Model Architecture}
The proposed X-SepFormer model is built on the backbone of SepFormer (a SOTA separation model). Given a speech mixture $y = x_{s} + x_{i}$, X-SepFormer aims to extract the target speech $x_s \in \mathcal{R}^T$ from the mixture $y\in \mathcal{R}^T$ while ignoring the interfering speech $x_i$, with additional reference speech $r_s \in \mathcal{R}^{T_r}$ (with duration $T_r << T$).  

The main architecture of X-SepFormer is illustrated in Fig.\ref{fig1}. Different from original SepFormer, our X-SepFormer increases the iteration number of the Intra-InterT processing from 2 to 4 and simplifies the Transformer encoder from 8 to 4 layers. The ECAPA-TDNN \cite{desplanques2020ecapa} based speaker embedding $e$ is extracted from $r_s$ and employed as the target speaker information. 

We found in preliminary experiments that the cross-attention-based speaker-speech fusion technique \cite{wang21} performed best and placing speaker embedding $e$ before the 1st layer of both Intra- and Inter-Transformer block achieved best performance. That results in combining $e$ with the speech mixture at 8 different depths of the SepFormer network. More details can be found in the later sections. 


\subsection{Model Training Objective} 
We expect our X-SepFormer have two key properties: (i) it attains good reconstruction quality of the extracted speech; and (ii) it alleviates the chunk-level SC errors. To this end, we firstly brief the existing training objective, then define the concept of chunk-level SC, finally propose two new training objectives.

\noindent\textbf{Original Training Objective}

The SI-SDR \cite{roux19} is commonly adopted in TSE to evaluate performance for speech enhancement systems.
Generally, it has a good correlation with speech quality. As such, most existing TSE models have been trained using the SI-SDR based loss function:$\mathcal{L_{\text{SI-SDR}}} = -\text{SI-SDR}$, with focus on the reconstruction quality of separated waveform.  

\noindent\textbf{Chunkwise SC}

In order to measure wrong speaker identify present in $\hat x$ at chunk level, system input $y, x_s$ and output $\hat x$ need to be split into sequence of small chunks, represented as $\{y^k\}_{k=1}^M$, $\{x_s^k\}_{k=1}^M$ and $\{\hat x^k\}_{k=1}^M$, respectively. Here the chunk number $M=\lceil (T-L)/O+1 \rceil$, with given chunk length $L$ and hop length $O$.
As aforementioned it has been reported \cite{zhang20m,elminshawi22} that negative SI-SDRi is a good approximation to the speaker extraction error rate, which also matches to our subjective observation. Thus we extend the idea to a chunkwise version, defined as:
$\text{SI-SDRi}^k = \text{SI-SDR}(\hat x^k, x_{s}^k)-\text{SI-SDR}(\hat x^k, y^k)$. 
If $\text{SI-SDRi}^k < 0$, we can infer the $k$-th chunk is identified with SC issue.

Based on above definition, we propose to statistically measure the chunkwise SC as the number ratio between identified SC chunks and total speech active chunks:
\begin{equation}
\begin{split}
    N_{sc} &= \sum_{k=1}^M \mathbb{I}(\text{SI-SNRi}^k <0) \\
    N_{valid} &= \sum_{k=1}^M \mathbb{I}(x_s^k>\eta)\cdot \mathbb{I}(\hat x^k>\eta) \\ \label{eta}
    r_{scr} &= N_{sc}/N_{valid}\times 100\%
\end{split}
\end{equation}
where $\mathbb{I}$ is the indicator function, $\eta$ denotes an energy-related threshold and $r_{scr}$ is the proposed measurement metric for the chunkwise SC.

\noindent\textbf{Proposed First Loss Scheme}

To address chunkwise SC issue, the motivation behind our first proposal is to scale the utterance-level reconstruction loss $\mathcal{L}_{\text{SI-SDR}}$, based on the statistic metric $r_{scr}$. We refer to it as scaled SI-SDR loss ($\mathcal{L}_{\text{scale-SI-SDR}}$), which is formulated as:
\begin{equation}
 \begin{aligned}
 &\mathcal{L}_{\text{scale-SI-SDR}} = -\alpha \cdot \text{SI-SDR} \\   \label{eqL1}
   \alpha &=\left\{
\begin{array}{rcl}
\gamma_1 - \gamma_2\cdot r_{scr},  &{\text{if SI-SDR} \geq 0}\\
\gamma_1 + \gamma_2\cdot r_{scr},  &\text{otherwise}
\end{array} \right.
\end{aligned}
\end{equation}

\noindent\textbf{Proposed Second Loss Scheme}

Alternatively, to adjust the $\mathcal{L}_{\text{SI-SDR}}$ more delicately, our second idea is to utilize the chunkwise information  $\text{SI-SDRi}^k$. However, in order not to distract the optimization of the TSE model, instead of directly use these values, we propose to explore the frequency distribution of sequence $\{\text{SI-SDRi}^k\}_{k=1}^M$. In particular, these values are grouped into 4 classes, with class intervals of $(-\infty,-5], (-5,0],(0,5],(5,\infty)$ respectively; and corresponding frequencies are denoted as $s_0,s_1,s_2,s_3$. Since groups with negative $\text{SI-SDRi}^k$ suggest speaker confused chunks, we further adopt weight argument. To sum up, as our second training objective, a weighted SI-SDR loss ($\mathcal{L}_{\text{weight-SI-SDR}}$) is given as:
\begin{equation}
    \mathcal{L}_{\text{weight-SI-SDR}} = -\frac{1}{N_{valid}}\sum_{j=0}^3 \omega_j\cdot s_j \label{eqL2}
\end{equation}
where weights follow the setting $\omega_0\geq \omega_1\geq \omega_2 \geq \omega_3>0$ to reflect class-dependent contributions to loss.

\section{Experiments and Analysis}
\subsection{Experiment Setup}
The WSJ0-2mix dataset is a popular speech separation dataset for two-speaker speech mixture. Based on it, like prior work, our TSE dataset is created following the script\footnotemark.

\footnotetext[1]{{https://github.com/xuchenglin28/speaker\_extraction\_SpEx/tree/master/\\data/wsj0\_2mix}}



Our X-SepFormer system is trained with batchsize of 1, using \textit{Adam} optimizer and following the learning rate scheduler of \textit{ReduceLROnPlateau}. The initial learning rate is set to $1.5e-4$ for the first 20 epochs then decreased to half when no improvement on the validation is observed over 2 epochs. Furthermore, our system training includes two stages. The system is firstly trained for 18 epochs using loss of $\mathcal{L}_{\text{SI-SDR}}$ to generate a base model; then different loss schemes are applied to finetune the base model to yield different TSE systems. The chunk parameters are set to $L$=250ms and $O$=125ms for training and $O$=0 for inference. We simply set $\gamma_1 = \gamma_2 = 1$ for $\mathcal{L}_{\text{scale-SI-SDR}}$ and weights $\omega_0 = \omega_1 = 5$, $\omega_2 = \omega_3 = 1$ for $\mathcal{L}_{\text{weight-SI-SDR}}$, to avoid the burden of careful hyper-parameter tuning. And the threshold in Eq.\ref{eta} is set to $\eta=15$.

Like most prior arts, we report SI-SDR and SI-SDRi for  evaluation of separation performance. In addition, to evaluate speech quality, both PESQ and proposed chunkwise SC metric ($r_{scr}$) are also reported. 
\subsection{Experiment Results}
\noindent\textbf{Proposed Systems vs. Baseline}

Firstly we compare three X-SepFormer systems, with the same system architecture but trained with different loss schemes. Comparison results are listed in Tab.\ref{tab:vsbase} and the best results among them are bold-faced.
\begin{table}[]
    \centering
    \resizebox{\linewidth}{!}
        {
        \begin{threeparttable}
        \begin{tabular}{c|c|c|c|c|c}
            \hline
            \hline
            System\tnote{$\dagger$} & training objective & SDRi & SI-SDRi & PESQ & $r_{scr}$ \\
            \hline
            $S_{base}$ &  $L_{\text{SI-SDR}}$ & 18.4 & 17.9 & 3.64 & 9.51\\ 
            \cline{1-6}
            $S_{sc}$ & $L_{\text{scale-SI-SDR}}$ & \textbf{18.8} & \textbf{18.3} & 3.66 & 9.03\\ 
            \cline{1-6}
            $S_{wt}$ & $L_{\text{weight-SI-SDR}}$ & \textbf{18.8} & \textbf{18.3} & \textbf{3.67} & \textbf{8.10} \\ 
            \cline{1-6} 
            \hline
            \hline
        \end{tabular}
        \begin{tablenotes}
            \item[$\dagger$] training with full utterance length
        \end{tablenotes}
        \end{threeparttable}
    }
    \caption{Effect of loss functions on X-SepFormer}
    \label{tab:vsbase}
\end{table}

From the results, it can be observed that the proposed baseline offers decent performances, showing less than one tenth chunks with SC issues. Both system $S_{sc}$ and $S_{wt}$ provide marginal improvement (by around 2\% relative) over the baseline in terms of SDRi, SI-SDRi and PESQ. However, on the specific metric of chunkwise SC ratio, substantial improvements (by 5.0\% and 14.8\% relative, respectively) are achieved with our proposed training objectives. These validate the effectiveness of both our proposals and have a positive impact on our subjective evaluation. Note that a similar performance trend is observed when adopting another type of speaker embedding, indicating such performance gains are not attributed to speaker embeddings. Lastly, comparing to system $S_{sc}$, $S_{wt}$ looks more promising by outperforming the baseline among all metrics.

\noindent\textbf{Proposed Systems vs. Prior Arts}

Now we compare our proposed TSE systems against top performing SOTA systems on the WSJ0-2mix. The overall comparison results are presented in Tab.\ref{tab:vsprior}. For a fair comparison, we cite the published results directly from literature and mark '-' if the corresponding result is not reported. 

\begin{table}[]
    \centering
    \resizebox{\linewidth}{!}{
    \begin{threeparttable}
    \begin{tabular}{l|c|c|c|c}
    \hline
    \hline
    Model & SDRi & SI-SDRi & PESQ & $r_{scr}$ \\ \hline
    Mixture & 0 & -0.001 & 2.01 & - \\ \hline
    Upper Bound\cite{lutati22} & - & 23.1 & - & - \\ \hline
    SepFormer(SS) & 22.4 & 22.3 & - & - \\ \hline
    SpEx\cite{chenglin20}\tnote{$\ast$} & 16.3 & 15.8 & - & -\\
    Spex+\cite{ge2020spex+}\tnote{$\ast$} & 17.2 & 16.9 & - & -\\
    Speakerbeam + DC\cite{katerina19} & 10.9 & - & - & - \\
    DPRNN-Spe-IRA\cite{luo2020dual}\tnote{$\ast$}  & 17.6 & 17.3 & 3.43 & - \\ 
    WASE\cite{hao2021wase}\tnote{$\ast$} & 17.0 & - & - & -\\ 
    SpExpc\cite{wang21}\tnote{$\ast$} & 18.8 & 18.6 & - & -\\
    SpExpc\cite{wang21}\tnote{$\ast$}  \tnote{$\dagger$} & 19.0 & 18.8 & - & -\\
    \hline
    X-SepFormer ($S_{base}$)\tnote{$\ast$}  & 19.5 & 18.9 & 3.74 & 9.17\\
    X-SepFormer ($S_{sc}$)\tnote{$\ast$} & \textbf{19.7} & \textbf{19.1} & \textbf{3.75} & 8.56\\
    X-SepFormer ($S_{wt}$)\tnote{$\ast$}  & 19.3 & 18.8 & 3.74 & \textbf{8.03}\\ \hline
    X-SepFormer ($S_{sc}$) + DA\tnote{$\ast$} & \textbf{20.1} & \textbf{19.5} & 3.80 & 8.47\\
    X-SepFormer ($S_{wt}$) + DA\tnote{$\ast$} & 19.9 & 19.4 & \textbf{3.81} & \textbf{7.14}\\\hline\hline
    \end{tabular}
    \begin{tablenotes}
    \item[$\ast$] training with truncated utterances (4s) instead of whole utterances  \end{tablenotes}
    \begin{tablenotes}
    \item[$\dagger$] adopting enrollment speech with fixed duration (60s) instead of average duration (7.3s)\end{tablenotes}
    \end{threeparttable}}
    \caption{Comparison of proposed systems with SOTA models on the WSJ0-2mix test set. (DA: data augmentation)}
    \label{tab:vsprior}
\end{table}

From the results in Tab.\ref{tab:vsprior}, we have the following key observations and conclusions:
\begin{enumerate}
    \item system $S_{base}$ is a strong baseline that outperforms those previously published results;
    \item three X-SepFormer systems ($S_{base}$, $S_{sc}$ and $S_{wt}$) are almost on a par with separation quality; while $S_{wt}$ performs best on chunkwise SC errors;
    \item data augmentation schemes (including dynamic mixture \cite{zeghidour21} and speed augmentation) contribute considerable performance improvements;
    \item our best performing system ($S_{sc}$+DA) delivers the best separation performance that narrows the gap between TSE and SS; 
    \item our best performing system ($S_{wt}$+DA) delivers the best SC performance;
    \item the relative performance gap between TSE and SS is reduced from 16.6\% to 12.6\%.
\end{enumerate}

\begin{figure}[htb]
\centering
\includegraphics[width=0.40\textwidth]{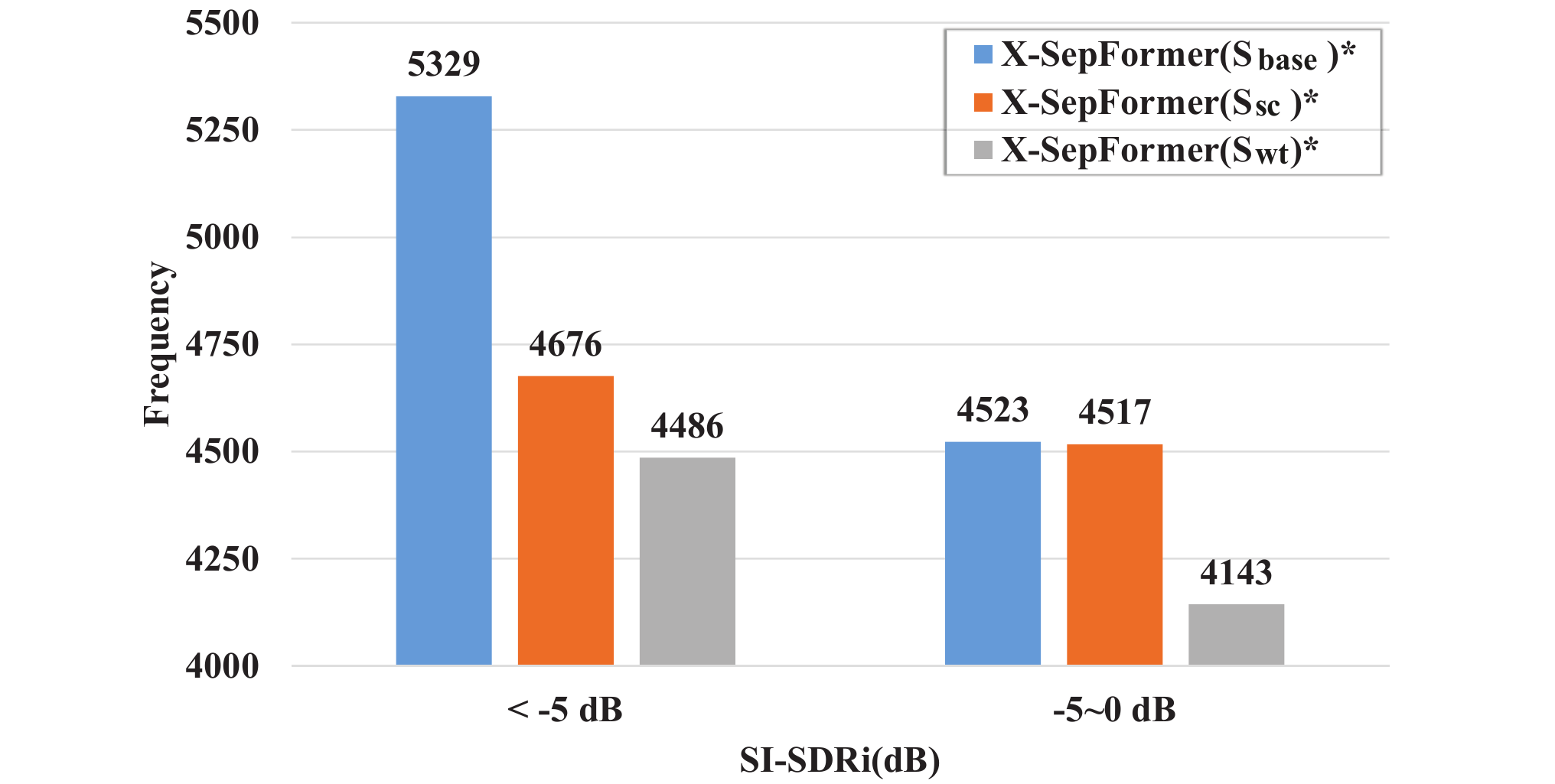}
\caption{Comparison of frequency distributions of speaker confused chunks}
\label{fig:fig2}
\end{figure}

Lastly, to highlight our proposals are advantageous on SC, Fig.\ref{fig:fig2} displays the frequency distributions of all SC chunks in extracted outputs on the WSJ0-2mix test set. 
Since herein we are interested in the SC chunks, only two clusters $\text{SI-SDRi}<-5\text{dB}$ and $-5 < \text{SI-SDRi}\leq 0\text{dB}$ are plotted in grouped bar plot. As expected, the numbers of SC chunks are reduced for both system $S_{sc}$ and $S_{wt}$. Corresponding audio samples are provided online for public assessment \footnotemark \footnotetext[2]{https://llearner.github.io/X-SepFormer.github.io/}.

\section{Conclusion}
In this paper, we introduced X-SepFormer: a TSE network architecture that is based on SepFormer with optimization objective on speaker confusion problem. For the first time, such a problem is quantitatively assessed via the metric of reconstruction quality improvement that is defined at small chunk-level. Two training schemes are proposed to leverage the quantitative metric and its associated distribution on extracted output. Our proposed training schemes are simple to implement, and add limited computational costs to the training process. In our experiments, both schemes are proven to be effective, and X-SepFormer even outperform recently proposed SOTA TSE systems in terms of both reconstruction quality and speech quality. Future research will extend our best system to other TSE datasets and evaluate it in real conversations.

\small
\bibliographystyle{unsrt}

\end{document}